\documentclass[final,3p,times,sort&compress,twocolumn]{elsarticle}

\usepackage{graphicx}

\usepackage{amssymb}
\usepackage{amsmath} 
\usepackage{epstopdf}
\usepackage{booktabs}
\usepackage{color} 
\usepackage{multirow}
\definecolor{gray}{rgb}{0.5, 0.5, 0.5}

\journal{Nuclear Instruments and Methods A}
\begin{document}
\begin{frontmatter}

\title{The Radon Monitoring System in Daya Bay Reactor Neutrino Experiment}

\author[cuhk]{M. C. Chu}
\author[cuhk]{K. K. Kwan}
\author[cuhk]{M. W. Kwok}
\author[hku]{T. Kwok}\ead{tnkwok@hku.hk}
\author[hku]{J. K. C. Leung}
\author[hku]{K. Y. Leung}
\author[cuhk,hku]{Y. C. Lin} 
\author[ucb,lbl]{K. B. Luk}
\author[hku]{C.~S.~J.~Pun}

\address[cuhk]{Department of Physics, The Chinese University of Hong Kong, Hong Kong, China}
\address[hku]{Department of Physics, The University of Hong Kong, Hong Kong, China}
\address[ucb]{Department of Physics, University of California at Berkeley, Berkeley, CA 94720, U.S.A.}
\address[lbl]{Physics Division, Lawrence Berkeley National Laboratory, Berkeley, CA 94720, U.S.A.}

\begin{abstract}
We developed a highly sensitive, reliable and portable automatic system (H$^{3}$) to monitor the radon concentration of the underground experimental halls of the Daya Bay Reactor Neutrino Experiment. H$^{3}$ is able to measure radon concentration with a statistical error less than 10\% in a 1-hour measurement of dehumidified air (R.H. 5\% at 25$^{\circ}$C) with radon concentration as low as 50 Bq/m$^{3}$. 
This is achieved by using a large radon progeny collection chamber, semiconductor $\alpha$-particle detector with high energy resolution, improved electronics and software. The integrated radon monitoring system is highly customizable to operate in different run modes at scheduled times and can be controlled remotely to sample radon in ambient air or in water from the water pools where the antineutrino detectors are being housed. The radon monitoring system has been running in the three experimental halls of the Daya Bay Reactor Neutrino Experiment since November 2013. 
\end{abstract}

\end{frontmatter}

\section{Introduction}\label{sec:1}
The main goal of the Daya Bay Reactor Neutrino Experiment is to determine the neutrino mixing angle $\theta_{13}$.
Eight anti-neutrino detector modules (ADs) are installed in three underground experimental halls. In order to suppress backgrounds related to cosmic-ray muons and natural radioactivity, the ADs are immersed in water pools \cite{bib:dybtdr}.

In an underground environment like this, radioactive radon gas can be easily found in air. 
Isotopes of Thorium and Uranium ( $^{232}$Th, $^{235}$U and $^{238}$U) in rocks produce gaseous $^{220}$Rn, $^{219}$Rn and $^{222}$Rn respectively in their decay chains. 
$^{219}$Rn in air has negligible impact to the experiment since its half-life is only 3.96~s. 
$^{220}$Rn has a half-life of 55.6~s, so it can exist for a longer time in air and may accumulate to a measurable quantity. 
For $^{222}$Rn, it has a half-life of 3.83 days. Numerous $\alpha$-particles are generated from $^{222}$Rn and its progenies along the chain: 
$^{222}_{86}$Rn 
$\xrightarrow[3.83 \rm{d}]{5.40 \> \rm{MeV} \> \alpha}$ 		$^{218}_{84}$Po
$\xrightarrow[3.05 \rm{m}]		{6.00 \> \rm{MeV} \> \alpha}$ 	$^{214}_{82}$Pb
$\xrightarrow[26.8 \rm{m}]		{\beta^{-}}$ 				$^{214}_{83}$Bi
$\xrightarrow[19.7 \rm{m}]		{\beta^{-}}$ 				$^{214}_{84}$Po
$\xrightarrow[164 \rm{\mu s}]	{7.69 \> \rm{MeV} \> \alpha}$ 	$^{210}_{82}$Pb
$\xrightarrow[22.3 \rm{y}]		{\beta^{-}}$ 				$^{210}_{83}$Bi
$\xrightarrow[5.01 \rm{d}]		{\beta^{-}}$ 				$^{210}_{84}$Po
$\xrightarrow[138 \rm{d}]		{5.30 \> \rm{MeV} \> \alpha}$	$^{206}_{82}$Pb.
These $\alpha$-particles can induce background in the experiment. Furthermore, due to its relatively long half-life, $^{222}$Rn can accumulate to a high concentration in poorly ventilated space and pose a health hazard. In view of its importance, the term ``radon" is used to refer specifically to $^{222}$Rn in this paper unless otherwise stated. 

Radon in air is particularly difficult to handle amongst common natural radioactive backgrounds. 
It can diffuse into the ADs through any unnoticeable leak or radon-permeable materials, and then dissolve in the liquid scintillator. The dissolved radon and its progenies can cause correlated background (from $\alpha$ decays) and singles background (from $\beta$ decays) mimicking antineutrino signals.
To reduce the amount of radon in the vicinity of the ADs, dry nitrogen gas is used to flush out air in 
the space between the water pool cover and the pool water, and in the dry ducts connected to the ADs. The pool water is also purified and recirculated. In addition to suppressing radon in the experiment, there is a need to monitor continuously its concentration, especially in the air in the experimental halls, and in the water surrounding the ADs.

A preliminary survey of airborne radon was carried out with an AB-5 detector (Pylon Electronics Inc.)~\cite{bib:ab-5} in the experimental halls (EH1, EH2, and EH3) of Daya Bay. Radon concentrations from about 50~Bq/m$^{3}$ to a few hundreds of Bq/m$^{3}$ were recorded at various locations. We require the statistical error of a meaningful monitoring of the radon concentration to be less than 10\% for 1-hour sampling of air with radon concentration down to 50 Bq/m$^{3}$. It is convenient to express this requirement by a detector calibration factor (C.F.) of 30~Bq/m$^{3}$/cpm with the definition:

\begin{equation}
\rm{C.F.} = \frac{\rm{radon ~concentration ~[Bq/m^{3}]} }{\rm{number ~of ~counts ~per ~minute ~[cpm]}}
\end{equation}
Also, the detector should be sensitive to a low radon concentration of 0.5 Bq/m$^{3}$ and portable, for carrying out \textit{ad hoc} radon measurements (e.g. radon in gas flushing the ADs).

The monitoring system should operate as an integrated system. It has to control the measurements of radon in air and water automatically following a customizable schedule. The radon data and the measurement conditions should be ready to be processed and uploaded to a centralized system of the experiment.

Thus a highly sensitive radon monitoring system was developed to meet these requirements. Details of this system are introduced in this paper. In Section \ref{sec:3}, the design and construction of the system are presented. Calibration and optimization of the radon detector are described in Section \ref{sec:4}. In Section \ref{sec:5}, the performance of the system in the underground halls is discussed.

\section{Other radon monitoring equipment}
Many commercial radon detectors employ Lucas cell or electrostatic method for radon progenies counting. Lucas cell-based detectors like AB-5 are renowned for their stability and simplicity of operation. However, the cell does not distinguish the $\alpha$-particles generated from radon, $^{218}$Po and $^{214}$Po. As the two nuclides preceding $^{214}$Po have half-lives of about 20 minutes, it takes about 3 hours for radon and $^{214}$Po to reach equilibrium inside the Lucas cell. This type of detector is not designed for monitoring the hourly fluctuation of radon continuously. 

Detectors employing electrostatic method to collect radon progenies can have a quicker response. Many of them are able to take an energy spectrum of $\alpha$-particles, enabling the selection of the fast-response $^{218}$Po events for counting. RAD7 (Durridge Company Inc.) \cite{bib:rad7} is a typical and popular model of this type of detector. However, due to the small volume of its internal sample cell (0.7 L), RAD7 has a C.F. of only 148~Bq/m$^{3}$/cpm. Another detector of this type is ERS-2-S (Tracerlab GmbH) \cite{bib:tracerlab}, which has a higher sensitivity. However, the $\alpha$-particle energy spectrum has limited resolution for easy background identification.

Excellent sensitivity reaching 1~mBq/m$^{3}$ or better can be achieved by radon extraction facilities built for some underground experiments \cite{bib:borexino, bib:superk, bib:sno}. These facilities utilize collection chamber of hundreds of litres in volume. They take up large spaces and are not intended to be portable.

\section{Radon monitoring in the Daya Bay Experiment}\label{sec:3}
In order to meet the requirements stated in Section \ref{sec:1}, the Hong Kong High-sensitivity High-reliability detector (H$^{3}$) is designed to have a C.F. better than 30~Bq/m$^{3}$/cpm with superb portability. The whole system is integrated to the hardware and software of the Daya Bay experiment in a simple manner. It is capable to control radon measurements in air and water samples, and present the real-time results through network.

\begin{figure*}[ht]
\begin{center}
\includegraphics[scale=0.6]{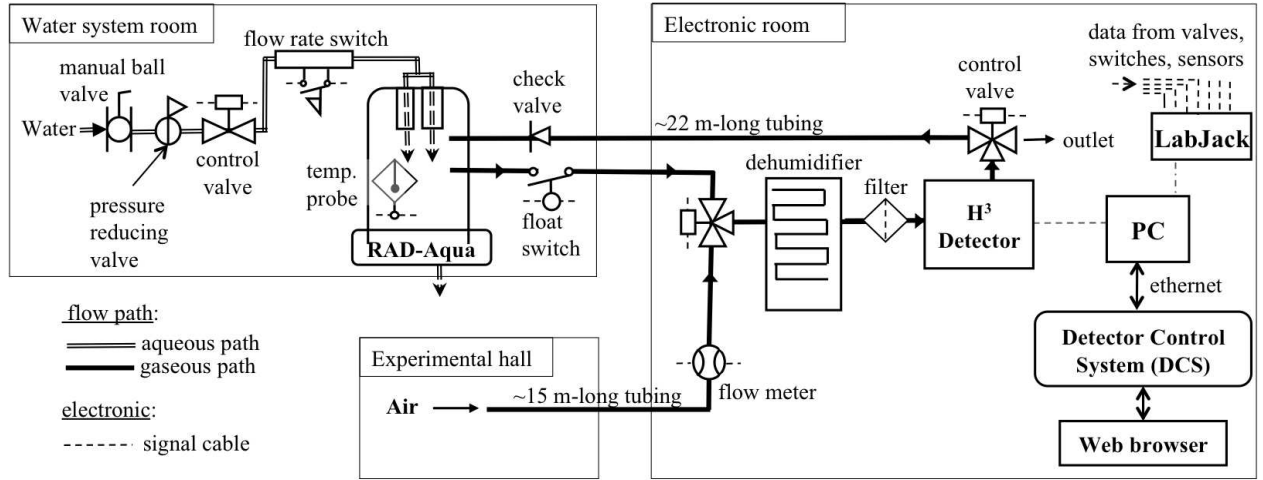} 
\caption{Schematic diagram of the radon monitoring system in each Daya Bay experimental hall. Details of components can be found in Sections \ref{sec:4} and \ref{sec:5}.}
\label{fig:system}
\end{center}
\end{figure*} 

As depicted in Figure~\ref{fig:system}, the system is installed in the electronics room of each experimental hall. There are two types of run mode: ``air run" for measuring the radon concentration
of the air sample extracted from the hall; and ``water run" for measuring the radon concentration of the water sample obtained from the recirculation system of the water pool. The samples are not required to flow at high speed or be pressurized. The sampling points can be set up easily by plugging the tubings into the push-in fittings.

The H$^{3}$ is designed to measure airborne radon directly. 
For radon dissolved in water sample, the gas is allowed to diffuse from water to air in a degassing unit first. Then the air is pumped to the H$^{3}$ for measurement. 

An software interface is written in LabVIEW Virtual Instrument (VI) \cite{bib:labview} to control and monitor the operation of the radon monitoring system. The host computer (host PC) of the LabVIEW VI is connected to a set of control valves and flow sensors installed along the sample pathways through a USB data acquisition module. 
The VI also reads the H$^{3}$ data via a RS-232 port. Then it processes the data with the calibration constants set by the system administrator. A copy of the processed data is also sent to the remote Detector Control System (DCS) of the
experiment which centralizes all the environmental data and detector operation conditions. 

\subsection{H$^{3}$ detector}\label{sec:3.1}
The H$^{3}$ detector is made of a rack-mount VME-4U chassis (178 mm (H) $\times$ 482 mm (W) $\times$ 279 mm (L)) that houses a radon 
progeny collection chamber, an $\alpha$-particle detector, a high-voltage generator, and front-end electronics, weighing less than 5 kg in total.

\subsubsection{Detection method}\label{sec:3.1.1}
Air sample is drawn into the radon progeny collection chamber for measurement. Radon progenies and dusts in the incoming air are removed by a glass fiber filter (GF-75, pore size 0.3 $\mu$m, Sterlitech Corp.)~\cite{bib:sterlitech} at the inlet of the chamber, so only the freshly produced radon progenies are collected.

In the chamber, the positively charged progeny $^{218}$Po are collected electrostatically on the entrance window of a Passivated Implanted Planar Silicon detector (PIPS model A450, Canberra Industries Inc.) \cite{bib:pips}. The PIPS measures
the energy of the $\alpha$-particles emitted in the decays of $^{218}$Po and subsequent progenies. Then the radon concentration can be calculated from the counts of $\alpha$-particle in the energy spectrum and the C.F.

\subsubsection{Radon progeny collection chamber}\label{sec:3.1.2}
\begin{figure}[h]
\begin{center}
\includegraphics[scale=0.62]{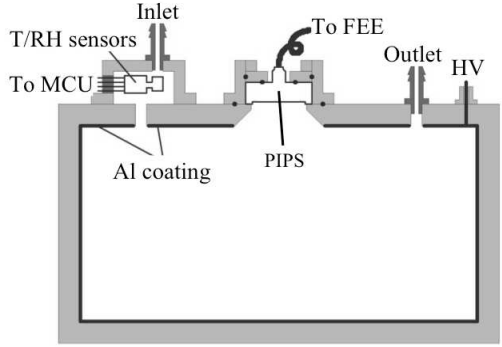} 
\caption{Cross section of the radon progeny collection chamber of H$^{3}$. O-rings used for ensuring air-tightness are depicted by the black dots around the PIPS.}
\label{fig:collection_chamber}
\end{center}
\end{figure}

The volume of the chamber required to achieve a C.F. of 30 Bq/m$^{3}$/cpm is estimated by:
\begin{equation}
\label{eqn:2}
V=\frac{1}{C.F. \times r_{+} \times \varepsilon_{p} \times \varepsilon_{\alpha} \times 60~\rm{[sec]} }
\end{equation}
where $r_{+}$ = 88\% \cite{bib:hopke} is the fraction of emanated $^{218}$Po that carries positive charge,
$\varepsilon_{p}$ is the collection efficiency of $^{218}$Po ion inside the chamber, and 
$\varepsilon_{\alpha}$ is the $\alpha$-particle detection efficiency of the PIPS. 
For collection chambers of similar mechanism and comparable size, $\varepsilon_{p}$ is about 70\% \cite{bib:wada}. 
For the PIPS, $\varepsilon_{\alpha}$ is 40\%. So the volume of the chamber of the H$^{3}$ should be about 2.25~L.
Figure~\ref{fig:collection_chamber} is the cross section of the radon progeny collection chamber. 
It is a polyvinyl chloride cylinder of 9-cm~(h) $\times$ 18-cm~($\varnothing$) (2.3 L) to provide sufficient volume of air sample to generate the progenies. 

The inner surfaces of the chamber are coated with aluminum foils. The PIPS is mounted on a circular aperture at the center of the top plate. Air sample is drawn into the chamber with a micro-pump installed downstream of the gas outlet. Temperature and relative humidity (R.H.) of the
air sample are monitored by digital sensors at the inlet of the chamber.

The aluminum coating of the chamber is connected to +1800 V, while the entrance window of the PIPS has a DC bias of +24 V. This establishes an electric field inside
the chamber for driving the positively charged radon progenies to the PIPS. 

The C.F.s obtained in the calibration (Section \ref{sec:4.4}) with this chamber were consistent with the estimation of Equation \ref{eqn:2}.
A larger chamber can boost the detector sensitivity as more radon progenies could be collected. However, this would sacrifice portability and require a highly stable high-voltage supply.

\subsubsection{Electronics unit}\label{sec:3.1.3}
\begin{figure}[h]
\begin{center}
\includegraphics[scale=0.3]{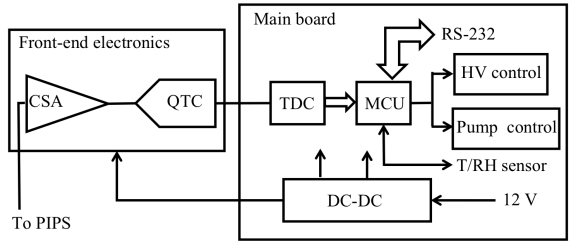} 
\caption{Electronics unit of the H$^{3}$. }
\label{fig:electronics_unit}
\end{center}
\end{figure}

Signals from the PIPS are analyzed by the on-board electronics unit. The electronics unit is divided into a front-end (FEE) circuit and a main board (Figure~\ref{fig:electronics_unit}).

The analog FEE circuit is shielded from the noisy digital and switching power circuits with a Faraday cage. 
It consists of a charge-sensitive amplifier (CSA, OP Amp AD8065) and a charge-to-time converter (QTC). The QTC converts an input analog pulse from the PIPS into a TTL logic signal with a width proportional to the input charge. Then the QTC output is digitized by the main board (32-bit ARM Micro Controller Unit).
The main board is also responsible for relaying commands from the host PC to different detector components (e.g. high voltage generator, pump).

\subsection{Degassing unit}\label{sec:3.2}
For water run, a RAD-Aqua (Durridge Company Inc.) \cite{bib:rad_aqua} is used as the degassing unit. The water sample is broken into droplets to facilitate radon diffusion from water to the air flowing through the RAD-Aqua. This flowing air circulates in a closed loop between the RAD-Aqua and the H$^{3}$. When diffusion equilibrium is reached for radon and with sufficient air exchange between H$^{3}$ and the RAD-Aqua, the air in the closed loop truly reflects the radon concentration in water.

\section{Performance and calibration of H$^{3}$}\label{sec:4}
In the designing stage, tests were carried out to optimize the performance of the H$^{3}$.

\subsection{Regions-of-interest for counting}\label{sec:4.1} 
\begin{figure}[h]
\begin{center}
\includegraphics[scale=0.48]{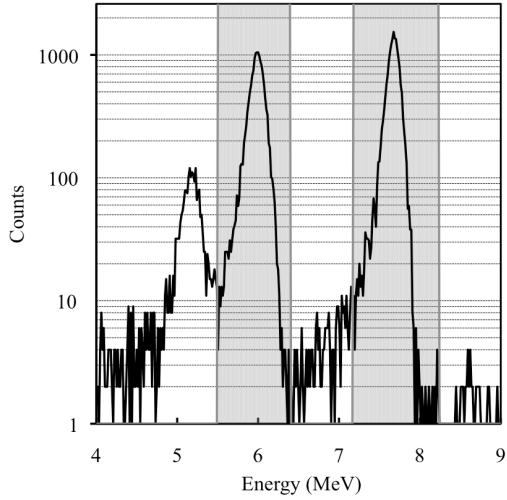}
\caption{Energy spectrum obtained with a H$^{3}$ detector. The grey areas are the ROI of $^{218}$Po (left) and $^{214}$Po (right).}
\label{fig:spectrum}
\end{center}
\end{figure}

Figure~\ref{fig:spectrum} shows the energy spectrum of  $\alpha$-particles collected by the H$^{3}$.
The two prominent peaks corresponding to the decays of $^{218}$Po and $^{214}$Po (6.00~MeV and 7.69~MeV respectively). In prolonged measurement, the $\alpha$-peak of $^{210}$Po (5.30~MeV) becomes visible gradually when $^{218}$Po and $^{214}$Po accumulate on the PIPS. However, the $^{210}$Po-peak does not respond quickly to any change of radon concentration. Thus it would not be used in the analysis.

The energy spectrum obtained by the H$^{3}$ has sufficient energy resolution for defining the regions-of-interest (ROIs) for $^{218}$Po and $^{214}$Po. 
The number of counts in the ROIs can be used to determine the radon concentration. The ROI of $^{218}$Po is 5.50 to 6.40 MeV, and 7.20 to 8.24~MeV for $^{214}$Po, each covering a range of five FWHMs around the respective peaks. Although results from both ROIs will be recorded by the H$^{3}$, only the counts in the $^{218}$Po peak are used to calculate the real-time radon concentration. This is because $^{218}$Po ($\tau_{1/2}$=3.05 min) is significantly quicker in response to the change of radon concentration than $^{214}$Po. 

Defining new ROI can enhance the identification and rejection of background or noise. 
Equation \ref{eqn:bi-212} shows the decays of the $^{220}$Rn progeny $^{212}$Bi that generates a 6.10-MeV $\alpha$-particle falling into the ROI of $^{218}$Po.

\begin{equation}
\label{eqn:bi-212}
\begin{aligned}
& ^{212}\rm{Bi} \xrightarrow[36\% \ , \ 60.6 \> \rm{m}]{6.10 \> \rm{MeV} \> \alpha}  	\ ^{208}\rm{Tl}	\\ 
& ^{212}\rm{Bi} \xrightarrow[64\% \ , \ 60.6 \> \rm{m}]{\ \ \  \ \ \  \beta^{+} \ \ \ \ \ } 						\ ^{212}\rm{Po}
\xrightarrow[299 \> \rm{ns}]{8.78 \> \rm{MeV} \> \alpha} 				\ ^{208}\rm{Pb}
\end{aligned}
\end{equation}
This background, which may only contribute up to a few percent of the total counts if the radon concentration is low and the sampling point of air is very close to the H$^{3}$, can be subtracted by adding an additional ROI from 8.24 to 9.26 MeV, to count the number of 8.78-MeV $\alpha$-particles associated with $^{212}$Bi. By multiplying this number by 36/64, the contamination of $^{212}$Bi decay in the ROI of $^{218}$Po can be calculated. 

\subsection{Calibration in exposure chamber}\label{sec:4.2}
\begin{figure}[h]
\begin{center}
\includegraphics[scale=0.25]{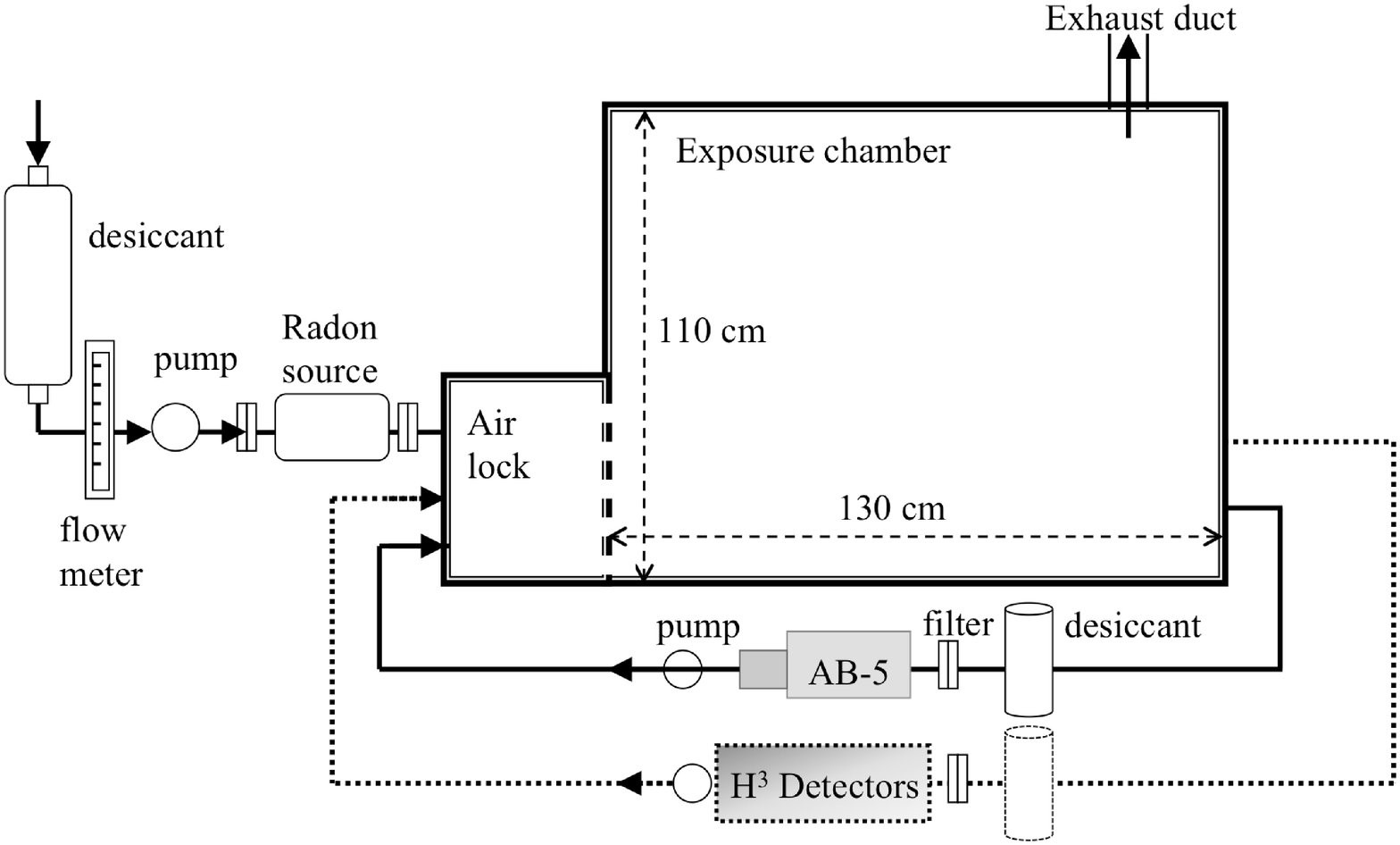} 
\caption{Schematic drawing of the exposure chamber for calibration in HKU.}
\label{fig:calib_setup}
\end{center}
\end{figure}

Calibration of the H$^{3}$ was carried out with a radon exposure chamber in The University of Hong Kong (HKU)~(Figure~\ref{fig:calib_setup}). The exposure chamber provided air samples of known and stable radon concentration with controlled environmental conditions \cite{bib:leung}. Air samples with radon concentrations from 200 Bq/m$^{3}$ to over 7000~Bq/m$^{3}$ with an uncertainty less than 5\% could be prepared.

Radon was added to the chamber by pumping air through a standard radon source (RN-1025, Pylon Electronics Inc.) \cite{bib:cal_source}. During H$^{3}$ calibration, the radon concentration was monitored by an AB-5 detector, whose calibration factor was traceable to the National Radon Laboratory of the University of South China in China. With the air samples of high radon concentration, sufficient statistics could be obtained in a short time for the calibration runs, such that any fluctuations caused by the environment could be identified easily by repeated measurements.

\subsection{Humidity}\label{sec:4.3}
\begin{figure}[h]
\begin{center}
\includegraphics[scale=0.52]{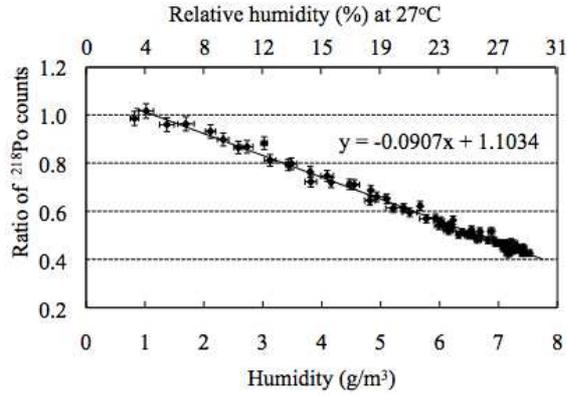} 
\caption{Ratio of $^{218}$Po counts from the H$^{3}$ being calibrated, to that from the control H$^{3}$. Horizontal axis shows the humidity of air sample in the H$^{3}$ being calibrated. Sample in the control H$^{3}$ was kept at 1.00$\pm$0.02 g/m$^{3}$. Statistical error (1-$\sigma$) was less than 2\%.}
\label{fig:diff_rh}
\end{center}
\end{figure}

Sensitivity of radon detectors using electrostatic collection method is inversely correlated to humidity~\cite{bib:tan}. Water vapor can neutralize the electric charge of the radon progenies, leading to a drop in the collection efficiency. Therefore, it is necessary to control and monitor the humidity of the air sample inside H$^{3}$ to maintain data integrity. During calibration, the air samples for the AB-5 and H$^{3}$ were dried with desiccant to a low R.H. of 5\% at 25$\pm$1$^{\circ}$C, which was equivalent to a water content of 1.15 g/m$^{3}$ in air. However, in real operation, humidity may deviate from this low value. So a correction curve for humidity was obtained for each H$^{3}$. This was done by connecting two H$^{3}$s independently to the radon chamber. One H$^{3}$ was used as a control, while the other one was under calibration. At the inlet of the control H$^{3}$, the humidity of the incoming air sample was kept at 1.00$\pm$0.02 g/m$^{3}$ throughout the calibration with a large column of desiccant. For the H$^{3}$ under test, the desiccant column was much smaller, such that the humidity was allowed to change slowly by less than 1.00~g/m$^{3}$ in 2 hours. Since each run lasted for 30 minutes only, the change in humidity of air sample in this H$^{3}$ was less than 0.25~g/m$^{3}$ and hence considered as stable. Besides humidity, both H$^{3}$s operated under the same conditions for over 30 hours. The ratio of the $^{218}$Po counts collected by the H$^{3}$s were plotted against humidity in Figure \ref{fig:diff_rh}. The number from the control H$^{3}$ was the denominator. The curve of each H$^{3}$ would be used to correct the C.F. for air sample of different humidity.

\subsection{Calibration factors of H$^{3}$}\label{sec:4.4}
Each H$^{3}$ was tested for determining the C.F.s of $^{218}$Po and $^{214}$Po as summarized in Table~\ref{table:1}. The test was repeated with a range of radon concentrations. No correlation of C.F.s with radon concentrations could be seen up to nearly 10$^{4}$ Bq/m$^{3}$.

\begin{table}[h]
\begin{center}
\begin{tabular}{c | c c}
\hline 
Radon conc. 			& \multirow{2}{*}{C.F. of $^{218}$Po}	& \multirow{2}{*}{C.F. of $^{214}$Po} \\	
by AB-5		\\		
(Bq/m$^{3}$) 			& (Bq/m$^{3}$/cpm) 		& (Bq/m$^{3}$/cpm)	\\
\hline \hline 
565.1 $\pm$ 4.3		& 27.51 $\pm$ 0.25 	 	& 25.72 $\pm$ 0.13 	 \\
795.2 $\pm$ 5.8 	 	& 26.60 $\pm$ 0.23 	 	& 24.93 $\pm$ 0.12 	 \\
1251 $\pm$ 8.3			& 28.13 $\pm$ 0.22 	 	& 25.76 $\pm$ 0.11 	 \\
7315 $\pm$ 21 	 		& 27.15 $\pm$ 0.97 	 	& 25.12 $\pm$ 0.10 	 \\
\hline \hline
Average C.F. 			& 27.35 $\pm$ 0.20		& 25.38 $\pm$ 0.10  \\
\hline
\end{tabular}
\caption{Calibration factors of a H$^{3}$ at different radon concentrations (R.H. = 5\%, temp = 25$\pm$1$^{\circ}$C). Statistical errors (1-$\sigma$) are shown for the C.F. at different radon concentrations. Standard errors are shown for the average C.F.}
\label{table:1}
\end{center}
\end{table}

\subsection{High-voltage optimization}\label{sec:4.5}
\begin{figure}[h]
\begin{center}
\includegraphics[scale=0.48]{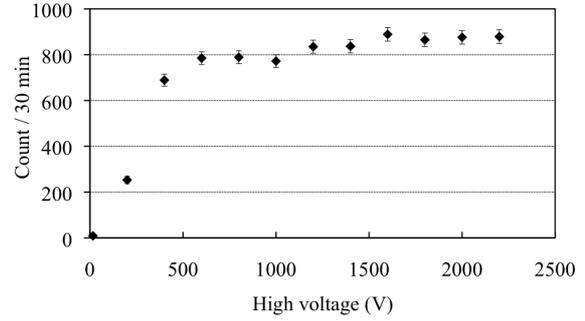} 
\caption{Number of counts of $^{218}$Po versus the HV applied to the radon progeny collection chamber. Radon concentration was 800$\pm$40 Bq/m$^{3}$ (R.H.=5\%, temp=25$\pm$1$^{\circ}$C).}
\label{fig:hv_cpm}
\end{center}
\end{figure}

The high voltage (HV) applied to the radon progeny collection chamber directly affects the collection efficiency $\varepsilon_{p}$ in Equation \ref{eqn:2}. A series of runs were conducted with different HV values to find out the optimal range. Air sample with a fixed radon concentration (800$\pm$40 Bq/m$^{3}$) was pumped from the exposure chamber to the H$^{3}$. The number of counts in the ROI of $^{218}$Po were plotted against the applied HV in Figure~\ref{fig:hv_cpm}. From the data, the number of counts began to plateau above 600 V. The optimal HV was chosen to be 1800~V, such that fluctuation of 1\% in the HV 
provided by the supply (DW-P202-IC33, DongWen High Voltage Power Supply Co.) would not lead to significant variation in the data.

\subsection{Pump operating at different flow rates}\label{sec:4.6}
\begin{figure}[h]
\begin{center}
\includegraphics[scale=0.51]{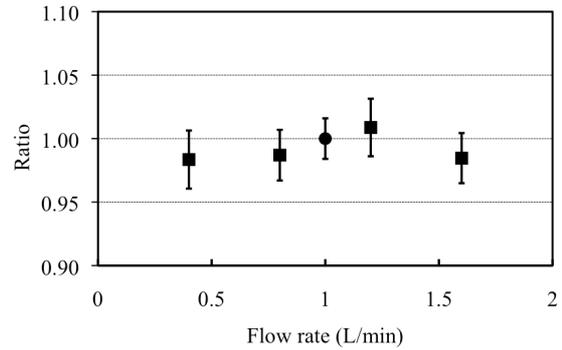} 
\caption{Ratio of average counts in the ROI of $^{218}$Po at different flow rates. Datum obtained at a flow rate of 1 L/min ($\bullet$) is the denominator. Radon concentration was 250$\pm$12 Bq/m$^{3}$.}
\label{fig:flowrate}
\end{center}
\end{figure}

The H$^{3}$ employs active and continuous sampling of air. Air sample is drawn into the collection chamber with a micro-pump at a flow rate of 1.00$\pm$0.05~L/min. Various flow rates between 0.4 to 1.6~L/min were also tested on the H$^{3}$. At each flow rate, the H$^{3}$ measured the air sample of stable radon concentration (250$\pm$12~Bq/m$^{3}$) for 2 hours. The number of counts in the ROI of $^{218}$Po are compared in Figure~\ref{fig:flowrate} by taking the 1-L/min datum as the denominator. In $\pm$20\% of the selected flow rate for operation (1~L/min), no systematic change in the ratio was observed.

\subsection{Background and lower limit of detection }\label{sec:4.7}
Background was measured by using radon-free nitrogen gas. Before the measurement, the H$^{3}$ was purged with nitrogen to remove any residual radon, and the remaining progenies in the collection chamber were allowed to decay. When the counts in the ROIs approached zero, the H$^{3}$ was run for 106 hours at a sample flow rate of 1.0$\pm$0.1 L/min. 

The average background rates were 0.68$\pm$0.08~counts/hr for $^{218}$Po, and 0.17$\pm$0.04 counts/hr for $^{214}$Po. They were equivalent to a radon concentration of 0.31$\pm$0.03~Bq/m$^{3}$ and 0.07$\pm$0.02~Bq/m$^{3}$ respectively by the C.F.s in Table~\ref{table:1}. The 90\% confidence level detection limit was 0.11~Bq/m$^{3}$. The background is sufficiently low that the H$^{3}$ can measure very low radon concentration for specific tasks, such as detecting radon in the AD cover gas.

\subsection{Precision of measurement}\label{sec:4.8}
\begin{figure}[h]
\begin{center}
\includegraphics[scale=0.52]{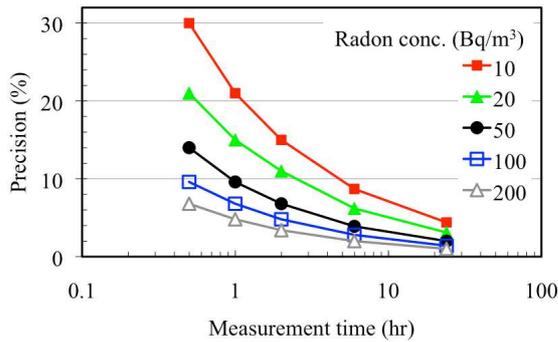} 
\caption{Precision of H$^{3}$ as a function of measuring time obtained by experiments with various radon concentrations (R.H.=5\%).}
\label{fig:precision}
\end{center}
\end{figure}

The precision of the H$^{3}$ is mainly affected by the statistical error of the counts of $\alpha$-particle. Figure \ref{fig:precision} summarizes the precision of the H$^{3}$ in a series of measurement using low radon concentrations, in which the counting statistics is low. The number of counts in the ROI of $^{218}$Po was used in the calculations. In the experimental halls at Daya Bay, the lowest radon concentration ever measured was about 50~Bq/m$^{3}$. At this radon concentration, uncertainty of 10\% or less can be achieved with an hour of measurement.

\subsection{Length of preparation period for water run} \label{sec:4.9}
Prior to data taking in a water run, a preparation period is needed for radon concentration in air and water to reach equilibrium in the degassing unit. Preparation time is also needed for the H$^{3}$ to reach a quiescent state after the previous air run.
As indicated in our preliminary survey, radon concentration in the air of the experiment hall is higher than that in the pool water by about a factor of six.
If the H$^{3}$ starts to collect data without the preparation period, the $\alpha$-particle energy spectrum would consist of considerable contribution of the radon progenies accumulated during the previous air run.

\begin{figure}[h]
\begin{center}
\includegraphics[scale=0.52]{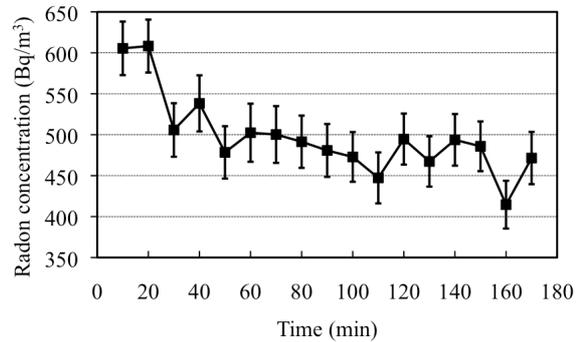}  
\caption{Results of H$^{3}$ measurement before and after switching on the RAD-Aqua at $t=$~20~minutes. Initial radon concentration of air from the exposure chamber was 606$\pm$45~Bq/m$^{3}$. Radon concentration in running water in RAD-Aqua was 106$\pm$6~Bq/m$^{3}$.}
\label{fig:water_eqm}
\end{center}
\end{figure}

To decide the length of the preparation period, the H$^{3}$ and RAD-Aqua were set up in HKU, resembling the details of the monitoring system. The H$^{3}$ was first flushed thoroughly with an air sample of radon concentration 606$\pm$45~Bq/m$^{3}$ from the exposure chamber. Then a water sample with 106$\pm$6~Bq/m$^{3}$ of radon was pumped to the RAD-Aqua at 3.4$\pm$0.3~L/min. Figure \ref{fig:water_eqm} shows the radon concentration measured by the H$^{3}$ before and after the RAD-Aqua was started at $t$~=~20~minutes. After $t$~=~60~minutes, the radon concentration in H$^{3}$ stabilized at 477$\pm$7~Bq/m$^{3}$, a concentration consistent with the Ostwald Coefficient (Section \ref{sec:5.2}). So the preparation period of water run should be at least 40 minutes to obtain the true radon concentration in water.

\section{On-site setup of the radon monitoring system}\label{sec:5}
The H$^{3}$ is capable of monitoring the radon concentration in air and water of each experimental hall at scheduled times. The details of on-site operation are described below.

\subsection{Radon concentration in air} \label{sec:5.1}
Air sample is extracted near the water pool in the experimental halls. The sampling points are 1 m above ground, and at least 2 m away from the walls. This reduces the background due to the $\alpha$-particles from the short-lived $^{220}$Rn (6.29 MeV), $^{216}$Po (6.78 MeV) and other progenies of $^{220}$Rn to a negligible level. No significant difference in radon concentration could be found near the water pool in the experiment halls because of the strong ventilation.

\subsection{Radon concentration in water}\label{sec:5.2}
Water sample is extracted from the recirculation system which pumps the water from the water pool to the purification system. The radon concentration of the water sample is identical to that in the water pool. Sample extraction rate to the RAD-Aqua is set to 2 to 4~L/min for a reasonable diffusion rate from water to air in a closed loop leading to the H${^3}$.
At diffusion equilibrium, the radon concentrations in the two media (water/air) are correlated by the Ostwald Coefficient, which has a value of 0.226 at 25$^{\circ}$C \cite{bib:solubility}. This gives extra sensitivity to the H$^{3}$ measurement of radon concentration in water. For 1 unit of change in the radon concentration in the incoming water sample, that would lead to over 1/0.226=4.42 units of change in the air inside the closed loop. The drawback of this method is that about 40 minutes is needed to reach equilibrium. Thus the total length of Òwater runÓ is set to 4 hours so radon data of more than 3 hours can be collected.

\subsection{Dehumidification}\label{sec:5.3}
The H$^{3}$ is designed for long-term remote operation. 
Desiccant is not suitable for continuous dehumidification as it requires frequent replacement.
Thus the system includes a custom-built dehumidifier for uninterrupted control of the humidity. 

Inside the dehumidifier, the air sample passes through a long zigzag path along an aluminum heat exchanger for drying the air by condensation. The heat exchanger is cooled by two thermoelectric Peltier modules. Their temperature is stabilized by a micro-controller (AT89C52) and temperature sensors. It takes 4 hours for them to reach the operating temperature. Afterwards, the R.H. can be maintained below 26\% for air sample between 18-26$^{\circ}$C. 

\subsection{On-site calibration with dehumidifier}\label{sec:5.4}
The H$^{3}$ is able to operate continuously with the built-in dehumidifier. Since the R.H. is held steady at about 26\% by the dehumidifier instead of 5\% as achieved at HKU, 
on-site calibration of the C.F. is necessary. This also serves as a standard annual check and update of the C.F.s of the H$^{3}$. Results of the annual calibration in 2013 by radon the air sample prepared by the exposure chamber are shown in Table \ref{table:onsite_cf}. Due to the difference in humidity, they were substantially different from those obtained with the exposure chamber in HKU (Table~\ref{table:1}). 
\begin{table}[h]
\begin{center}
\begin{tabular}{c | c c c}
\hline 
				 & EH1 & EH2 & EH3 \\
\hline \hline
Mean temp ($^{\circ}$C) & 26.9$\pm$0.2 & 27.5$\pm$0.3 & 25.9$\pm$0.3 \\
R.H. (\%) & 20.0$\pm$0.2 & 24.7$\pm$0.3 & 25.3$\pm$0.3 \\
\hline
\multicolumn{4}{c}{}\\
\hline
Progeny & \multicolumn{3}{c}{C.F. (Bq/m$^{3}$/cpm)} \\
\hline \hline
$^{218}$Po & 39.3$\pm$0.7 & 37.1$\pm$0.7 & 43.5$\pm$1.5 \\
$^{214}$Po & 35.2$\pm$0.6 & 35.5$\pm$0.7 & 39.5$\pm$1.6 \\
\hline
\end{tabular}
\caption{On-site calibration factors of the H$^{3}$s with dehumidifiers. Statistical errors (1-$\sigma$) are shown.}
\label{table:onsite_cf}
\end{center}
\end{table}

\subsection{Control and data-taking software}\label{sec:5.5}
\begin{figure}[ht]
\begin{center}
\includegraphics[scale=0.5]{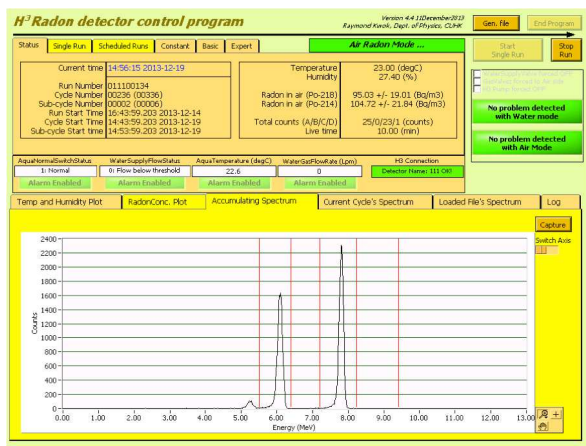} 
\caption{H$^{3}$ control interface on the host PC.}
\label{fig:interface}
\end{center}
\end{figure}

Figure~\ref{fig:interface} shows the LabVIEW VI interface developed for managing the two types of run of H$^{3}$. 
Mode-switching is automatic, following a user-defined schedule set in the VI. All running parameters and standard values of sensors for different operations are stored in a VI library.
A daily measurement schedule includes an air run (20 hours), and a water run (3 hours) preceded by a preparation period (1 hour). At the beginning of an air run, the VI would set the valve positions so that the H$^{3}$ would be flushed with air in the experimental hall. In the preparation period of a water run, the RAD-Aqua would be switched on so diffusion equilibrium can be reached by the time measurement starts.

Through a LabJack USB data acquisition module (U3-LV, LabJack Corporation, USA), status of the sensors and controllers are monitored by the VI. Several contingent modes are programmed in the VI.
An alarm would be triggered if any reading drifts outside the predefined operation range. Based on the category of abnormal reading, the VI would run the corresponding preset contingent mode or halt the erroneous operation. 
E-mail alert would be sent to the experts.
A full history of these readings and contingencies would be recorded by the host PC for diagnosis. 

The VI reads and processes the energy spectrum of $\alpha$-particles, and plots the H$^{3}$ internal temperature, humidity, and values of all sensors and control valves. The results are stored in a local hard-disk and simultaneously copied to the network database of DCS. 

\subsection{Radon concentration in AD cover gas}\label{sec:5.6}
For each AD, a flexible duct for keeping the detector cables and gas line dry is sealed off with o-rings. To remove contaminants and radon that permeates through the o-rings, the duct is constantly flushed with radon-free boil-off nitrogen (R.H.$\ll$5\%) from the AD cover gas system \cite{bib:dyb_gas_sys}. 
The returning gas from each AD was measured with the H$^{3}$ to ensure the radon concentration in the duct was under control. 

A different configuration was used to measure the radon concentration in the AD cover gas.
To prevent the relatively radon-rich air in the experimental hall from contaminating the air sample, the low-level measurements were carried out with a H$^{3}$ sealed inside an acrylic box. The returning gas flowed through the H$^{3}$ directly at 1 L/min before exhausting into the acrylic box. At the outlet of the acrylic box, back-diffusion from the experimental hall was cut off
with a mineral oil bubbler operating at a positive pressure of 2 cm of water. 
The radon concentration was calculated using the C.F.s in Table \ref{table:1}, which are applicable to the cover gas with R.H.$<$5\%. Typically, the radon concentration in the cover gas was in the sub-Bq/m${^3}$ level. 

\subsection{Uncertainty budget}\label{sec:5.7}
Besides the statistical uncertainty (maximum of 10\% at 1-$\sigma$), Table \ref{table:uncertainty_budget} summarizes other uncertainties of the radon concentration calculated from the $^{218}$Po counts of H$^{3}$. 
The most significant one is the 6.10\% from the C.F.~($^{218}$Po) of H$^{3}$. It is the quadratic sum of the uncertainty of radon concentration in the exposure chamber (5\% maximum), and the maximum standard error of the repeated measurement of the C.F. (3.5\% at EH3 in Table \ref{table:onsite_cf}). 
This uncertainty of C.F. already includes the uncertainties caused by the pump and the HV of H$^{3}$.
For the uncertainty induced by the pump operating at different flow rates, no systematic trend in the performance could be observed at $\pm$20\% of 1~L/min.
Furthermore, the pump is able to operate stably within 5\% at the designated flow rate of 1 L/min. Thus the upper bound of uncertainty it causes is taken to be the 1.3\%, which is the statistical uncertainty at 1~L/min in Figure \ref{fig:flowrate}.

For the HV, its fluctuation only have a small contribution of 0.08\% as evaluated from the data in Figure \ref{fig:hv_cpm}. The uncertainty is small because the H$^{3}$ operates in the plateau region of HV. 

The humidity correction applied to the C.F. also inherits the errors in the fit of Figure \ref{fig:diff_rh}, which is 1.26\%.

Uncertainty caused by the operation of the PIPS is negligible.  
Any tiny shift in the energy scale can be compensated by the flexible definition of the ROIs.
The energy resolution of the PIPS has been very stable since the commissioning of the radon monitoring system. 

\begin{table*}[h]
\begin{center}
\noindent
\begin{tabular}{l cccc}
\hline 
Quantity				&	Absolute					& Relative 		& Uncertainty of	\\
					&							& (\%) 			& radon conc. (\%) 	\\
\hline \hline
C.F. ($^{218}$Po)		&							& 				&	6.1	\\
$\>\>$ - Exposure chamber &							& 5						\\
$\>\>$ - Repeatability (incl. flow rate, HV)	&				& 3.5						\\
Humidity corr.			&							& 				&	1.26	\\
$\>\>$ - slope			&	0.001					& 1.13			&		\\
$\>\>$ - y-intercept		&	0.006					& 0.57			&		\\
\hline \hline
Overall				&							&				&	6.2	\\
\hline
\end{tabular}
\caption{Summary of uncertainties other than counting statistics in the ROIs.}
\label{table:uncertainty_budget}
\end{center}
\end{table*}

\section{Results}\label{sec:6}
Installation of the radon monitoring system on-site was completed in 2013. The 
data from the H$^{3}$s are uploaded to the DCS of the Daya Bay Experiment. 
Users can access the history of radon concentration and other parameters 
(temperature, R.H., etc.) through the DCS.  

\subsection{Radon in air on-site}\label{sec:6.1}
The plots in the left column of Figure~\ref{fig:eh_air_water} show the radon concentration in air since November 2013 for the three experimental halls at Daya Bay. The daily and monthly averages are plotted. It shows that the radon concentrations are quite stable. The H$^{3}$ also recorded spikes in the hourly radon concentration of nearly 900~Bq/m$^{3}$ in EH2, and 400~Bq/m$^{3}$ in EH3, when the ventilation systems of those halls were shut down for maintenance in January 2014.

\subsection{Radon in water on-site}\label{sec:6.2}
\begin{figure*}[h]
\begin{center}
\includegraphics[scale=0.31]{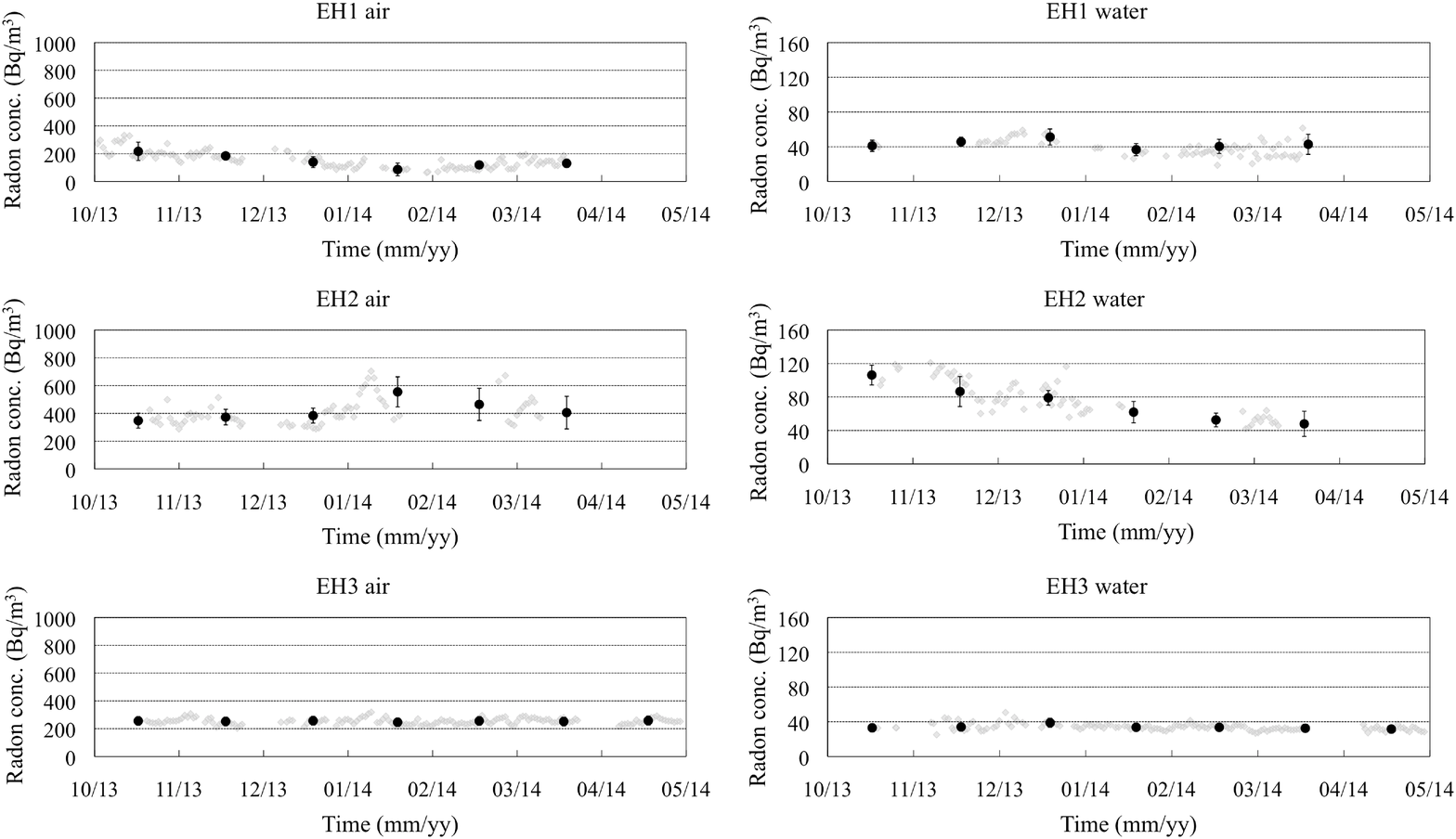}
\caption{Measured radon concentration in air (left), and in water for the water pool (right) of the Daya Bay experimental halls. Daily average is denoted by {\color{gray}$\blacklozenge$} and monthly average by $\bullet$. }
\label{fig:eh_air_water}
\end{center}
\end{figure*}

The automatic radon monitors for the water pools were installed in November 2013. The plots in the right column of Figure~\ref{fig:eh_air_water} show the daily and monthly averaged radon concentration in water for the water pools of the Daya Bay Experiment. The measured radon concentrations are consistent with the results of 34.7$\pm$3.5 to 86.4$\pm$8.6 Bq/m$^{3}$ obtained in person before the installation of the H$^{3}$s. The greatest advantage of the automation is that the radon concentration in water can be monitored continuously and remotely. So far the radon concentrations in the water were stable.

\subsection{Cover gas of Daya Bay ADs}\label{sec:6.3}
For the six ADs installed in 2012, the radon concentrations were found to be below 0.55 Bq/m$^{3}$, over 400 times lower than the ambient environment (Table~\ref{table:4}). The H$^{3}$ results showed that the AD cover gas system was operating as designed and was able to limit radon from accumulating in the ADs to an insignificant level.
\begin{table*}[h]
\begin{center}
\begin{tabular}{c | c | c | c  | c | c | c}
\hline 
						&  \multicolumn{2}{c|}{EH1}		& EH2 		&  \multicolumn{3}{c}{EH3} \\
\hline
Ambient radon (Bq/m$^{3}$)	& \multicolumn{2}{c|}{136$\pm$44 } 	& 221$\pm$40 &  \multicolumn{3}{c}{260$\pm$40} \\
\hline 
						& AD1 		& AD2 		& AD3 		& AD4 		& AD5 		& AD6 \\
\hline \hline
Integrated time (min)			& 370 		& 890 		& 905 		& 905 		& 530 		& 895 \\
\hline
Counts ($^{218}$Po)		& 2 			& 10 			& 6 			& 12 			& 6 			& 9 \\
Counts ($^{214}$Po)		& 0 			& 1 			& 8 			& 4 			& 5 			& 3 \\		
\hline  \hline
Radon conc. (Bq/m$^{3}$)	& $<$0.34		& $<$0.50 	& $<$0.38 	& $<$0.22 	& $<$0.55		 & $<$0.45 \\	
\hline
\end{tabular}
\caption{Radon concentration (95\% C.L.) of returning gas (R.H.$<$5\%) of the AD cover gas systems of Daya Bay in 2012 \cite{bib:dyb_gas_sys}.}
\label{table:4}
\end{center}
\end{table*}

\section{Conclusion}\label{sec:7}
We have developed a radon monitoring system for the Daya Bay Reactor Neutrino Experiment that has been running routinely for over two years.  
It can measure radon concentration either in air or in water 
with a precision significantly better than most commercial units in a relative short time, fulfilling our needs of rapid monitoring the presence
of radon which potentially can generate background in the experiment.   
In addition, with a custom-designed data acquisition system, we can
remotely control the continuous operation of the otherwise automatic monitoring system, thus significantly reducing the demand of human resource. 

\section{Acknowledgement}
We are grateful for the support with grants from the Research Grant Council of the Hong Kong Special Administrative Region, China (Project Nos. CUHK 1/07C and CUHK3/CRF/10) and from the University of Hong Kong (Project code: 201007176191).
K.B.L. is supported by the Office of Science, Office of High Energy Physics, 
of the U.S. Department of Energy under Contract No.~DE-AC02-05CH11231.
 \\

\end{document}